\documentstyle[12pt]{article}
\newlength{\myleftmargin}
\newlength{\paperwidth}
\begin{document}
 \rightline{\bf KOBE-FHD-94-09}
 \rightline{\bf September 1994}
 \vskip 0.5 cm
\renewcommand{\thefootnote}{\fnsymbol{footnote}}
\begin{center}
THE SPIN STRUCTURE OF NUCLEONS AND \\
DEEP INELASTIC SCATTERINGS\footnote[1]{Invited talk presented at the XII
International Seminar on High Energy Physics Problems-Relativistic Nuclear
Physics and Quantum Chromodynamics-, Dubna, 12-17 September, 1994.} \\

\parskip 1em

\underline{T. Morii} \\

\parskip 1em

Faculty of Human Development, Kobe University \\
Kobe 657, Japan, E-mail addresss: morii@jpnyitp.bitnet

\parskip 1em

{\bf Abstract}
\end{center}

\noindent
Based on a simple model which is compatible with the idea of
the static quark model and the parton model,
the polarized structure functions of proton and deuteron,
two-spin asymmetries of $\pi ^0$ in polarized $pp$ reactions and
inelastic $J/\psi$ productions in polarized lepton-proton collisions
are analyzed.
In particular, an important role of polarized gluon distributions is
pointed out.

\parskip 1em

\noindent
{\bf 1. Introduction}

The advent of {\it so--called} ``the proton spin crisis''
which has emerged from
the measurement of $g_1^p(x)$ by the EMC Collaboration[1], has stimulated a
great theoretical and experimental activity in particle physics[2].
So far various theoretical approaches have been provided to get rid of the
crisis.
Although some of them are very successful, a lot of problems remain to be
solved.
The problem is still very challenging topics in particle physics.
In this Talk, after briefly reviewing
what the problem is,
I would like to discuss the physics of spin effects in
various processes from a rather conservative point of view,
{\it i.e.} based on a simple model which is compatible with the idea of
the naive quark model and parton model. Furthermore, polarized
gluon distributions are examined in detail.

\noindent
{\bf 2. Proton spin problem}

In the kinematical region where the one-photon exchange is dominant,
the differential cross section of the deep inelastic
lepton-nucleon scattering,
$\ell +{\rm N}\rightarrow \ell'+X$,
is given by the product of
the lepton tensor $L^{\mu\nu}$ and the hadron tensor $W_{\mu\nu}$.
The antisymmetric part of $W_{\mu\nu}$ under $\mu\leftrightarrow\nu$
is described as
\begin{equation}
W_{\mu\nu}^{(A)}=\epsilon_{\mu\nu\rho\sigma}q^{\rho}\left[
s^{\sigma}\left\{M_NG_1(\nu, q^2)+\frac{p\cdot q}{M_N}G_2(\nu, q^2)\right\}-
s\cdot qp^{\sigma}\frac{G_2(\nu, q^2)}{M_N}\right]~,
\end{equation}
where $G_{1,~2}$ are called the spin-dependent structure
functions. In terms of $G_{1,~2}$, the difference of differential
cross sections
$d\sigma_{\uparrow\uparrow}$ and $d\sigma_{\uparrow\downarrow}$, where the
helicities of the longitudinally polarized beam and target are parallel and
antiparallel, respectively, can be written as
\begin{equation}
\frac{d^2\Delta\sigma}{d\Omega dE'}
 = \frac{4\alpha^2E'}{EQ^2}\left\{(E+E'\cos\theta)M_NG_1(\nu, Q^2)-
 Q^2G_2(\nu, Q^2)\right\}~,
\end{equation}
where $\theta$ is the lepton scattering angle in the lab.
frame, and $E$ and $E'$ are
the initial and final lepton energies, respectively.
$G_2$ in eq.(2) is suppressed
with respect to $G_1$ by a factor $\frac{Q^2}{EM_N}\sim 0.01$, for a typical
beam energy of $100$ GeV.
In the region of deep inelastic scattering(DIS),
$G_{1,~2}$ have a scaling property(Bjorken scaling);
\begin{equation}
M_N^2\nu G_1(\nu, Q^2)  \rightarrow   g_1(x)~,~~~~~~
M_N\nu^2G_2(\nu, Q^2)\ \rightarrow \ g_2(x)~,
\end{equation}
where $x=\frac{Q^2}{2M_N\nu}$ is a dimensionless scaling variable.
The Bjorken scaling has been understood
well by the parton model:  the DIS is viewed as an incoherent sum of
elastic scatterings of leptons by pointlike constituents inside a nucleon.
According to the parton model, $g_1(x)$ is described as
\begin{equation}
g_1(x)=\frac{1}{2}\sum_ie_i^2\left\{q_{i~\uparrow}(x)-q_{i~\downarrow}(x)+
 \bar q_{i~\uparrow}(x)-\bar q_{i~\downarrow}(x)\right\}
 \equiv \frac{1}{2}\sum_ie_i^2\delta q_i(x)~,
\end{equation}
where the sum is taken over the various species of partons with charge $e_i$
($i$=u, d, s, c, $\cdots$). $q_{i~\uparrow}(x)$ ($q_{i~\downarrow}(x)$)
represents the parton distribution polarized in parallel (antiparallel) to the
nucleon spin with the momentum fraction $x$ of the nucleon.
It is well known that the Bjorken scaling is violated even at large $Q^2$.
Tis is due to anomalous dimensions of the flavor singlet composite operators
appearing in the operator product expansion of the electromagnetic current
and running of the strong coupling constant $\alpha_S$.
Perturbative QCD describes well such scaling violations.

In 1988, EMC group reported[1]
\begin{eqnarray}
\int^1_0g_1^p(x, Q^2)dx&=&\frac{1}{2}\int^1_0\left\{\frac{4}{9}\delta u(x, Q^2)
+\frac{1}{9}\delta d(x, Q^2)+\frac{1}{9}\delta s(x, Q^2)\right\}dx
\nonumber\\
&=&\frac{1}{2}\left\{\frac{4}{9}\Delta u(Q^2)+\frac{1}{9}\Delta d(Q^2)+
\frac{1}{9}\Delta s(Q^2)\right\}\\
&=&0.126\pm 0.010 (stat.)\pm 0.015
(syst.)~.
\end{eqnarray}
at $Q^2=10.7$ GeV$^2$, where
$\frac{1}{2}\Delta q_i(Q^2) = \frac{1}{2}\int_0^1 \delta q_i(x,Q^2)dx$
means the spin
carried by quark $i$ in the proton.  By combining the data on
neutron $\beta$ decays,
$\Delta u-\Delta d=F+D=1.259\pm 0.006$[3]
and hyperons $\beta$ decays,
$\Delta u+\Delta d-2\Delta s=3F-D=0.688\pm 0.035$[4],
they obtained
$\frac{1}{2}\Delta u=0.391\pm0.016\pm 0.023$,
$\frac{1}{2}\Delta d=-0.236\pm0.016\pm 0.023$ and
$\frac{1}{2}\Delta s=-0.095\pm0.016\pm 0.023$.
Then the sum of the quark spin contributions to the proton becomes
\begin{eqnarray}
\frac{1}{2}\Delta\Sigma=\frac{1}{2}\sum_i\Delta q_i&=&
\frac{1}{2}\left\{\Delta u+\Delta d+\Delta s\right\}\nonumber\\
&=&0.060\pm 0.047\pm 0.069~.
\end{eqnarray}
This implies that very little of the proton spin is carried by
quarks. Furthermore, the rather large $\Delta s$ is surprising.
The results are very different from the prediction by the static
quark model and also the Ellis-Jaffe sum rule[5] derived from current
algebra and the assumption of $\Delta s=0$. It
is called ``proton spin crisis".\footnote[2]{Very recently, SMC group reported
a little larger value of
the first moment of $g^{p}_1(x)$, $i.e.~\int_0^1{g^{p}_1(x)dx}
=0.136\pm 0.011(stat.)\pm 0.011(syst.)$ but the
main conclusion remains unchanged[6].}
So far a number of ideas have been proposed to get rid of the crisis.
Among them, there has been an interesting idea that gluons contribute
significantly to the proton spin through the $U_A(1)$ anomaly of QCD[7].
In this model, spin-dependent quark distributions are largely affected
by gluons and the amount of the proton spin carried by quarks is not
necessarily small, where the integrated value of the polarized gluons
($\Delta G(Q^2)$) in the proton becomes $5\sim 6$ at $Q^2=10.7GeV^2$.
In the next section, to get into deeper understanding of the problem,
I propose a different but simple model which is
based on rather conventional idea.

\noindent
{\bf 3. Model of spin-dependent distribution functions}

In the quark-parton model
a proton is composed of three valence quarks accompanied
by sea quarks and gluons, though it
consists of three constituent quarks alone in the static quark model.
As a working model of a proton which is compatible with these pictures,
we propose a new wavefunction of a proton described by a
superposition of three-,five-, $\cdots$, body wavefunctions of
quarks. In practice, $\grave a$ la Carlitz and Kaur[8] we consider that
a proton is composed of an active quark interacting with virtual
photon in $\ell p$ reactions and a remaining ``core''. Then,
a polarized proton wavefuncton is given by[9]
\begin{eqnarray}
\mid p \uparrow \rangle &=& a_0 \Big[\mid \Psi_0 \rangle
 + \mid \Psi_1 \rangle \Big]_V \nonumber \\
 &+& a_1 \Big[ \mid
   \Psi '_0 \rangle + \mid \Psi '_1 \rangle +{\frac{\epsilon}{\sqrt 2}}\big(
   \mid \Psi''_0 \rangle + \mid \Psi''_1 \rangle +\mid \Psi''_{\frac{1}{2}}
   \rangle\big) \Big]_{V+S}+\cdots,
\end{eqnarray}
where $V$ and $V+S$ mean that the constituents are valence quarks, $u_Vu_Vd_V$,
and valence plus sea quarks, $u_Vu_Vd_Vq_S\overline{q}_S$, respectively.
The suffix of $\Psi, \Psi'$ and $\Psi''$ represents the isospin of the
``core'', which is composed of $qq$ for $\Psi$ and $qqqq/ qqq \overline{q}$
for $\Psi'$ and $\Psi''$. $\mid \Psi'_0 \rangle$ and $\mid \Psi'_1 \rangle$
are constructed by $uudu\overline{u}$ and $uudd\overline{d}$.
Each $\mid \Psi''_i \rangle$ in the $\epsilon$ term
comes from $uuds\overline{s}$.
$a_0(a_1)$ is the weight of the three-
(five-) quark wavefunction. $\epsilon$ denotes the relative weight of
an $s$-quark pair to $u/d$-quark pairs.
The values of $a_0^2, a_1^2$ and $\epsilon^2$ are determined to be $1,~0.1425$
and $0.5$, respectively, so as to reproduce the magnetic moment[3] and the
$K/ \pi$ production ratio in hadron collisions[10].

Then, the spin-dependent distribution functions
of quarks
can be derived as follows[9]:
\begin{eqnarray}
 -\frac{2}{3}d_V(x)\rbrace +a_1^2\lbrace\frac{17}{12}u_V(x)
 -\frac{5}{4}d_V(x)\nonumber\\
 &+&\frac{17}{6}\overline{u}_S(x)-\frac{5}{2}\overline{d}_S(x)+
 \frac{\epsilon^2}{2}u_V(x)-\frac{\epsilon^2}{2}\frac{2}{3}d_V(x)
 \rbrace \Big],\nonumber\\
 \delta d(x) &=& \widehat{D_f}(x)\Big[a_0^2 \lbrace -\frac{1}{3}
 d_V(x)\rbrace +a_1^2\lbrace\frac{7}{12}u_V(x)-\frac{3}{4}d_V(x)
 \\
  &+&\frac{7}{6}\overline{u}_S(x)-\frac{3}{2}\overline{d}_S(x)-
  \frac{\epsilon^2}{2}\frac{1}{3}d_V(x)\rbrace \Big],\nonumber\\
 \delta s(x) &=& \widehat{D_f}(x)\Big[a_1^2\frac{\epsilon^2}{2}
 \frac{1}{3}\lbrace s_S(x)+\overline{s}_S(x)\rbrace \Big],\nonumber
\end{eqnarray}
with
$\widehat{D_f}(x) = \frac{D_f(x)}{a_0^2+a_1^2+{\frac{\epsilon^2}{2}}
a_1^2}$,
where $D_f(x)$ is called a spin-dilution factor introduced originally
in CK model[8] and measures the deviation of spin-dependent
distributions from the $SU(6)$ limit.
With $a_0=1$ and $a_1=0$, eq.(9) reduces to CK model as expected.
In general, $D_f(x, Q^2)$ can be written by
\begin{equation}
 D_f(x, Q^2) =\frac{\lbrace 1-2P_f(x)\rbrace H_0 N(x, Q^2) + 1}
 {H_0 N(x, Q^2) + 1},
\end{equation}
where $N(x,Q^2)$ is the density of gluons relative to the quarks and
$H_0$ is fixed to be $H_0=0.0055$ by
the Bjorken sum rule[11].
$P_f(x)$ is the probability of the quark spin flip due to interactions
between quarks and gluons, and is given as
$P_f(x) = \frac{\sigma_{\downarrow\uparrow}(x)}{\sigma_{\uparrow\uparrow}(x)
+ \sigma_{\downarrow\uparrow}(x)}$ as a function of $x$ by using the analogy
of Rutherford scattering[9]. With
$P_f(x)=\frac{1}{2}$, eq.(10) results in the one in CK model.
By using the Duke-Owens
parametrization for spin-independent
quark and gluon distribution functions[12], one can calculate the
spin-dependent quark distributions.
Furthermore, the effect of gluons on the first moment of $g_1^p(x)$ is taken
into account through the $U_A(1)$ anomaly .
Since at present we have no definite knowledge of the
polarized gluon distribution
functions, we simply assume\footnote[3]{In practice, eq.(11) has been taken
under some theoretical considerations
and numerical analyses[13].}(Fig.1)
\begin{equation}
\delta G(x, Q^2=10.7GeV^2) = Cx^{0.1}(1-x)^{17},
\end{equation}
where $C=3.1$ is determined so as to fit $\int_0^1{g^{p}_1(x)dx}
= 0.126$(EMC).
By taking eq.(11), the spin-dependent quark distributions are
modified from $\delta q$ to
$\widetilde{\delta q_i}(x, Q^2) = \delta q_i(x, Q^2)-
\frac{\alpha_S(Q^2)}{2\pi}
\delta G(x, Q^2)$. (Fig.1)
With the help of the results in Fig.1,
we can reproduce
the $x$ dependence of $g_1^p(x, Q^2_{EMC})$ [1] and
$g_1^d(x, Q^2_{SMC})$ [14]
(Figs.2 and 3).
Moreover, the model leads to $\Delta u = 1.002,
\Delta d = -0.256, \Delta s$ $= 0.019$ and hence
$\frac{1}{2}\left\{\Delta u + \Delta d + \Delta s\right\}
= 0.382$,
that is, $76\%$ of the proton spin is to be carried by quarks.
Note that the model predicts rather small $\Delta s$.
Owing to this small $\Delta s$,
the $U_A(1)$ anomaly inevitably leads to large gluon
polarizations$(\Delta G=6.32)$ in order to explain the EMC data.
However, is the gluon polarization really so large in the proton?
To confirm this result, it is absolutely necessary to measure, in experiment,
the physical quantity sensitive to polarized gluon distributions.
In the following section, gluon polarization
effects on various reactions are studied.

\noindent
{\bf 4. Way to probe polarized gluon distributions}

In this section, we are concentrated on
two interesting processes which give us
important informations on the polarized gluons: one is the
$\pi^0$ production in polarized proton-polarized
proton collisions and the other is the $J/\psi$ production in
polarized electron-polarized proton collisions.
Before getting into the discussion of these processes, I present
some typical examples of polarized gluon distributions considered here:
\noindent

(a) the present model~;
\begin{equation}
x\delta G(x, Q^2=10.7{\rm GeV}^2)=3.1x^{0.1}(1-x)^{17}
{}~{\rm with}\,~~
\Delta G(Q^2_{EMC})=6.32~,~~~~~~~
\label{eqn:typeA}
\end{equation}

(b) Cheng--Lai type model[15];
\begin{eqnarray}
&& x\delta G(x, Q^2=10{\rm GeV}^2)=3.34x^{0.31}(1-x)^{5.06}
(1-0.177x)~~~~~~~~~~~~~~~~~~~~\,~~~~~~~~~~~~\nonumber\\
&& {\rm with}~~
\Delta G(Q^2_{EMC})=5.64~,
\label{eqn:typeB}
\end{eqnarray}

(c) BBS model[16];
\begin{eqnarray}
&&x\delta G(x, Q^2=4{\rm GeV}^2)=0.281\left\{(1-x)^4-(1-x)^6\right\}+
1.1739\left\{(1-x)^5-(1-x)^7\right\} \nonumber\\
&&{\rm with}~~~\Delta G(Q^2_{EMC})=0.53~,
\label{eqn:typeC}
\end{eqnarray}

(d) no gluon polarization model;
\begin{equation}
x\delta G(x, Q^2=10{\rm GeV}^2)=0~~{\rm with}~~\,\Delta G(Q^2_{EMC})
=0~.~~~~~~~~~~~~~~~~~~~~~~~~~~~~~~~~
\label{eqn:typeD}
\end{equation}
Among these examples, $\Delta G$ of types (a) and (b) are large while those of
types (c) and (d) are small and zero, respectively.
The $x$ dependence of $x\delta G(x, Q^2)$ and $\delta G(x, Q^2)/G(x, Q^2)$
which are evolved up to $Q^2=10.7$ GeV$^2$ by the Altarelli--Parisi equations
are depicted in Fig.4 (A) and (B), respectively.
As for the $x\delta G(x, Q^2)$ with large $\Delta G$,
many people[17] have taken up so far the one similar to type (b).
As shown in Fig.4,
the $x\delta G(x)$ of type (b) has a peak at $x\approx 0.05$
and gradually decreases with increasing $x$ while that of (a) has a sharp peak
at $x<0.01$ and rapidly decreases.
The type (c) which is derived from the requirements of the color coherence at
$x\sim 0$ and the counting rule at $x\sim 1$ has no sharp peak
but distributes rather broadly.

\noindent
{\bf 4. 1. Two-spin asymmetry for $\pi ^0$ productions
in polarized $pp$ collisions}

The interesting physical parameter to be discussed
here is the two-spin asymmetry
$A_{LL}$ as a function of transverse momenta $p_T$ of produced particles like
$\pi^0$, $\gamma$ and J/$\psi$. $A_{LL}$ is defined as
\begin{eqnarray}
A_{LL}~&=&~\frac{\left[d\sigma_{\uparrow\uparrow}-d\sigma_{\uparrow\downarrow}+
d\sigma_{\downarrow\downarrow}-d\sigma_{\downarrow\uparrow}\right]}
{\left[d\sigma_{\uparrow\uparrow}+d\sigma_{\uparrow\downarrow}+
d\sigma_{\downarrow\downarrow}+d\sigma_{\downarrow\uparrow}\right]}
\nonumber\\
&=&~\frac{Ed\Delta\sigma/d^3p}{Ed\sigma/d^3p}~,
\label{eqn:dfnALL}
\end{eqnarray}
where $d\sigma_{\uparrow\downarrow}$, for instance, denotes that the helicity
of a beam particle is positive and that of a target particle is negative.
So far, $A_{LL}$ for only inclusive $\pi^0$-production has been measured
by the E581/704 Collaboration at Fermilab[18] by using longitudinally
polarized proton (antiproton) beams and longitudinally polarized proton
targets.
Two--spin asymmetries $A^{\pi^0}_{LL}(\stackrel{\scriptscriptstyle(-)}{p}
\stackrel{}{p})$ contain contributions of various subprocesses. The
difference between $A_{LL}^{\pi^0}(pp)$ and $A_{LL}^{\pi^0}(\overline{p}p)$
for theoretical calculations is due to the magnitude and sign of contributing
subprocesses to $pp$ and $\overline{p}p$ reactions. For subprocesses concerned
here, an incident $\overline{q}$ is a sea component for a proton while it is a
valence component for an antiproton. Hence, $q\overline{q}\rightarrow q
\overline{q}$, $q\overline{q}\rightarrow gg$ and $\overline{q}g\rightarrow
\overline{q}g$ contribute more to $\overline{p}p$ than to $pp$
reactions. On the other hand, $qq\rightarrow qq$ and $qg\rightarrow qg$
contribute more to $pp$ than to $\overline{p}p$ reactions.
Furthermore, the spin--dependent subprocess cross section
$d\Delta\hat \sigma/d\hat t$ is
negative for $q_i\overline{q}_i\rightarrow \overline{q}_iq_i$, $q_i
\overline{q}_i\rightarrow q_j\overline{q}_j/\overline{q}_jq_j$, $q_i
\overline{q}_i\rightarrow gg$ and $gg\rightarrow q_i\overline{q}_i$, while it
is positive for other subprocesses. Therefore, the spin--dependent differential
cross section $Ed\Delta\sigma/d^3p$ for $\overline{p}p$ reactions becomes a
little smaller than the one for $pp$ reactions. This leads to smaller $A_{LL}^{
\pi^0}(\overline{p}p)$ than $A_{LL}^{\pi^0}(pp)$ as shown in Figs.5 and 6.
Several people have analized these interesting data[19].
By comparing the data with the calculations by Ramsey and Sivers[19],
the E581/704 group has concluded that the large $\Delta G$
in the proton is ruled out[18].

Here by using the spin--dependent gluon distribution functions
((a)$\sim$(d)) presented above, we have calculated $A_{LL}^{\pi^0}(pp)$ and
$A_{LL}^{\pi^0}(\overline{p}p)$, which are shown in Figs.5 and 6 for
$\sqrt s=20$ GeV and $\theta =90^{\circ}$, respectively,
where we typically choose $Q^2=4p_T^2$ with the transverse momentum $p_T$ of
$\pi^0$.
Comparing theoretical predictions with the experimental data,
one can see that not
only the no gluon polarization model (type (d)) but also
the present model (type (a))
seem to be consistent with the experimental data for both
$pp$ and $\overline{p}p$ collisions. It is remarkable to see that type (a)
works well in spite of large $\Delta G$.
Owing to the kinematical constraint of $x$ in the hard--scattering parton
model, the contributions from $0<x<0.05$
to $A^{\pi^0}_{LL}(\stackrel{\scriptscriptstyle(-)}{p}\stackrel{}{p})$ are
vanishing. Accordingly, there are no significant contributions from the
spin--dependent gluon distribution of type (a) to $A_{LL}^{\pi^0}$ though
$\Delta G(Q^2)$ for this case is quite large.
However, if we take the polarized gluon distribution $x\delta G(x)$ of
type (b) which is still large for $x>0.05$, we have a significant contribution
from the large $x\delta G(x)$ to $A_{LL}^{\pi^0}$ and then the result
becomes inconsistent with the E581/704 data. Furthermore, if the value of
$x\delta G(x)$ is not very small for $x>0.15$ even though $\Delta G(x)$ is
small (as in the case of type (c)), the calculation might not agree with the
experimental data. Therefore, one can conclude that a
large gluon polarization inside a proton is not necessarily ruled out
but the shape of the spin--dependent gluon distribution function is
strongly constrained by the E581/704 data.

\noindent
{\bf 4. 2. $J/\psi$ productions in polarized $lp$ collisions}

As can be seen from the above analyses, one cannot distinguish types (a)
and (d) of $x\delta G$, as long as we remain in the
analysis on $A_{LL}^{\pi^0}$.
Here, to see more clearly the effect of the spin--dependent gluon
distributions, we study the J/$\psi$ production processes in polarized
$\ell$p collisions, which may serve as the most straightforward method for
extracting $\delta G$[20,21].
The difference of  types (a) and  (d) can be found from the analysis of
inelastic J/$\psi$ productions in polarized {\it ep} collisions[21].
In the inelastic region where the J/$\psi$ particles are produced via the
photon--gluon fusion, $\gamma^* g \rightarrow J/\psi ~g$,
the spin--dependent differential cross section is given by
\begin{equation}
\frac{d\Delta\sigma}{dx} = x\delta G(x, Q^2) \delta f(x, x_{min})~,
\label{eqn:ddsdx}
\end{equation}
where $\delta G(x, Q^2)$ is the spin--dependent gluon distribution
function and $x$ the fraction of the proton momentum carried by the initial
state gluon.  $\delta f$ is a function which is
sharply peaked at $x$ just above $x_{min}$ and given by[21]
\begin{eqnarray}
\delta f(x, x_{min})&=&\frac{16\pi\alpha_S^2\Gamma_{ee}}{3\alpha m_{J/\psi}^3}
\frac{x_{min}^2}{x^2} \label{eqn:df}\\
&\times&\left[\frac{x-x_{min}}{(x+x_{min})^2}+
\frac{2x_{min}x\ln\frac{x}{x_{min}}}{(x+x_{min})^3}-
\frac{x+x_{min}}{x(x-x_{min})}+
\frac{2x_{min}\ln\frac{x}{x_{min}}}{(x-x_{min})^2}\right]~,\nonumber
\end{eqnarray}
where $x_{min}\equiv m_{J/\psi}^2/s_T$ and $\sqrt {s_T}$ is the total energy
in photon--proton collisions.
Fig.7 shows the $x$ dependence of $d\Delta\sigma/dx$ calculated with types
of (a) and (b) for various energies including relevant HERA energies.
As $\delta f$ has a sharp peak, the observed cross section
$d\Delta\sigma/dx$ directly reflects the spin--dependent gluon distribution
near $x_{peak}$. As is seen from eq.(17), $d\Delta\sigma/dx$ is
linearly dependent on the spin--dependent gluon distribution. Thus, if
$\delta G(x)$ is small or vanishing, $d\Delta\sigma/dx$ must be necessarily
small.
We are eager for the result given in Fig.7 being checked in the forthcoming
experiments.

\noindent
{\bf 5. Discussion}


Before closing this Talk, I would like to give some comments on the remaining
problems.  One comment is on
the polarized
$s$ quarks. The EMC data suggest a large and negative
contribution of $s$ quarks to the proton spin, $\Delta s=-0.19$.
However, contrary to such a large $\Delta s$,
the experimental data on charm productions
in neutrino DIS gave a restrictive bound
$|\Delta s| \leq 0.057^{+0.023}_{-0.057}$[22]
which is in little agreement with the EMC results. A way to get rid of this
inconsistency might come from the U$_A(1)$ anomaly. If the U$_A(1)$ anomaly is
taken into consideration and $\Delta s$ from the EMC data is replaced by
$\widetilde{\Delta s}$, then
these data might be reconciled with each other by taking rather large
$\Delta G$. To confirm this
interpretation, one need to measure independently
the magnitude of both the polarized gluons and strange quarks.

Another comment is on the proton spin sum rule,
$\frac{1}{2}=\frac{1}{2}\Delta\Sigma+\Delta G+\langle L_Z\rangle_{q+G}$.
If $\Delta G$ is large ($\simeq 5\sim 6$),
we are to have an approximate
relation $\langle L_Z\rangle_{q+G}\simeq -\Delta G$.
However, at present nobody knows the underlying physics of what it means.
It remains to be a problem, though the idea of the U$_A(1)$ anomaly
is attractive.

The running and future experiments on spin physics by deep
inelastic scattrings are decisively important for
going  beyond the present
understanding on the hadron structure.

\noindent
Fig.~1~:~Modified spin--dependent distribution functions
of quarks and the spin--dependent distribution function of
gluons at $Q^2=10.7~{\rm GeV}^2$.

\parskip 1em

\noindent
Fig.~2~:~Comparison of the spin--dependent structure function
$xg{_1^p}(x,Q^2 = 10.7$ GeV$^2$) with expermental data.
The solid and the dashed lines denote the result of the present model and
the EMC fit, respectively.
The full circle, open triangle and square points show the EMC, SLAC (E80)
and SLAC(E130) data, respectively.
Inner and outer error bars mean the statistical and total errors,
respectively

\parskip 1em

\noindent
Fig.~3~:~The $x$ dependence of the spin-- dependent deuteron structure
function $xg_1^d(x, Q^2)$ at $Q^2=4.6$ GeV$^2$.
Experimental data are taken from[14].

\parskip 1em

\noindent
Fig.~4~:~The $x$ dependence of (A) $x\delta G(x$, $Q^2)$ and (B)
$\delta G(x, Q^2)/G(x, Q^2)$ for various types (a)--(d) given by
eqs.(12), (13), (14) and (15)
at $Q^2=10.7$ GeV$^2$.

\parskip 1em

\noindent
Fig.~5~:~Two--spin asymmetry $A_{LL}^{\pi^0}(pp)$ for $\sqrt s=20$ GeV
and $\theta=90^{\circ}$, calculated with various types of
$x\delta G(x)$, as a function of transverse momenta $p_T$ of $\pi^0$.
The solid, dashed, small--dashed and dash--dot-
ted lines indicate the results
using types (a), (b), (c) and (d) in eqs.(12), (13),
(14) and (15), respectively. Experimental data are
taken from[18].

\parskip 1em

\newpage
\noindent
Fig.~6~:~Two--spin asymmetry $A_{LL}^{\pi^0}(\overline{p}p)$ for
$\sqrt s=20$ GeV and $\theta=90^{\circ}$, calculated with types (a), (b), (c)
and (d) for $x\delta G(x)$, as a function of transverse
momenta $p_T$ of $\pi^0$. Data are taken from[18].

\parskip 1em

\noindent
Fig.~7~:~The distribution $d\Delta\sigma/dx$ predicted by using types
(a) and (b) of $x\delta G(x, Q^2)$, as a function of $x$ for different values
of $\sqrt{s_T}$. The solid (dashed) curve corresponds to type (a) (type (b)).

\end{document}